\documentclass[twocolumn, twocolappendix]{aastex63}

\shorttitle{Caught in swallowtails}
\shortauthors{A. K. Meena et al.}

\usepackage{txfonts}
\usepackage[T1]{fontenc}
\usepackage{longtable}
\usepackage{multirow}

\usepackage{hyperref}
\defcitealias{2014MNRAS.445..201M}{M14}
\defcitealias{2021MNRAS.508.1206J}{J21}

\usepackage{xspace}
\def\Sref#1{Sec.~\ref{#1}\xspace}
\def\Fref#1{Fig.~\ref{#1}\xspace}

\def\Tref#1{Table~\ref{#1}\xspace}

\begin{document}

\title[Caught in swallowtails]{Caught in  Swallowtails: \\ 
Discovery of Two Swallowtail Image Formations in MS~0451.6--0305}

\correspondingauthor{Ashish Kumar Meena}
\email{akm@iisc.ac.in}

\author[0000-0002-7876-4321]{Ashish Kumar Meena}
\affiliation{Department of Physics, Indian Institute of Science, Bengaluru 560012, India}

\author[0000-0003-1060-0723]{Wenlei Chen}
\affiliation{Department of Physics, Oklahoma State University, 145 Physical Sciences Building, Stillwater, OK 74078, USA}

\author[0000-0001-6278-032X]{Lukas J. Furtak}
\affiliation{Department of Astronomy, The University of Texas at Austin, Austin, TX 78712, USA}
\affiliation{Cosmic Frontier Center, The University of Texas at Austin, Austin, TX 78712, USA}

\author[0000-0001-5492-1049]{Johan Richard}
\affiliation{Univ Lyon, Univ Lyon1, Ens de Lyon, CNRS, Centre de Recherche Astrophysique de Lyon UMR5574, 69230 Saint-Genis-Laval, France}

\author[0000-0002-0350-4488]{Adi Zitrin}
\affiliation{Department of Physics, Ben-Gurion University of the Negev, P.O. Box 653, Beer-Sheva 8410501, Israel}

\author[0000-0001-9065-3926]{Jose M. Diego}
\affiliation{Instituto de F\'isica de Cantabria (CSIC-UC). Avda. Los Castros s/n. 39005 Santander, Spain}

\author[0000-0003-1974-8732]{Mathilde Jauzac}
\affiliation{Centre for Astrophysics Research, Department of Physics, Astronomy and Mathematics, University of Hertfordshire, Hatfield AL10 9AB, UK}
\affiliation{Centre for Extragalactic Astronomy, Durham University, South Road, Durham DH1 3LE, UK}
\affiliation{Institute for Computational Cosmology, Durham University, South Road, Durham DH1 3LE, UK}
\affiliation{Astrophysics Research Centre, University of KwaZulu-Natal, Westville Campus, Durban 4041, South Africa}

\author[0000-0003-3142-997X]{Patrick L. Kelly}
\affiliation{School of Physics \& Astronomy, University of Minnesota, 116 Church St. SE, Minneapolis, MN 55455}

\author[0000-0001-8156-6281]{Rogier A. Windhorst}
\affiliation{School of Earth and Space Exploration, Arizona State University, Tempe, AZ 85287-6004, USA}

\begin{abstract}
We report the discovery of two swallowtail image formations at~$z=2.91$ and~$z=6.70$ behind the galaxy cluster MS~0451.6--0305 in JWST-NIRCam imaging. We find that in both of the above lensed systems, the complex image morphology cannot be reproduced by simple fold/cusp caustics, and detailed lens modeling reveals higher-order swallowtail caustic configurations. In the~$z=2.91$ lens system, a small part of the source galaxy (which itself is part of a galaxy group) containing atleast two compact knots sits inside the swallowtail caustic, producing a quadruply imaged arc. At two of the image positions of these knots, we infer point source magnifications of~$\gtrsim 300$, implying lensing-corrected effective radii of~$\lesssim 0.8-1.5$~pc. The~$z=6.70$ system exhibits even more complex image formation. We therefore only use the most confidently identified counter-images of knots in this system as constraints in our lens modeling. The resulting model predicts magnifications~$\sim20-200$ and lensing-corrected effective radii of~$\lesssim 0.8-18.5$~pc for various knots. Together, these two systems represent the first example of observations of multiple swallowtail image formations in a single galaxy cluster and demonstrate the ability of swallowtail caustics to magnify individual substructures at sub-parsec scales, from intermediate redshifts to the first billion years of the Universe.
\end{abstract}


\section{Introduction}
\label{sec:intro}
Gravitational lensing by galaxy clusters with the James Webb Space Telescope~(JWST) has been a treasure trove, leading us to observations of lensed supernovae~\citep[e.g.,][]{2025ApJ...979...13P, 2025arXiv250912319S, 2026arXiv260104156C}, lensed stars~\citep[e.g.,][]{2022ApJ...940L..54C, 2023A&A...672A...3D, 2023ApJ...944L...6M, 2025NatAs...9..428F}, and small parsec scale details in distant galaxies~\citep[e.g.,][]{2022A&A...659A...2V, 2023ApJ...945...53V, 2024Natur.632..513A, 2025A&A...694A..59M, 2025arXiv250907073V, 2025arXiv251208054A}. Observation of small-scale structures in a background source occurs when the source lies close to a caustic, which traces high-magnification regions in the source plane, and the corresponding image formations appear close to critical curves. Caustics can be divided into different types, exhibiting distinct magnification distributions and image multiplicities in their vicinity~\citep[e.g.,][]{1992grle.book.....S, 2001stgl.book.....P}. Most of the highly magnified sources have been observed close to the \textit{fold} caustic, leading to a pair of bright images in the sky (excluding global minimum image). Although observationally rare~(one in every five to ten clusters;~\citealt{2021MNRAS.503.2097M}), there exist caustic structures, such as the swallowtail and (hyperbolic and elliptic)~umbilics, that can lead to four (or more) highly magnified images in a characteristic configuration, all of which lie close to each other in the sky~\citep[e.g.,][]{2020MNRAS.492.3294M, 2022MNRAS.515.4151M}.

\begin{figure*}
    \centering
    \includegraphics[width=1.0\textwidth]{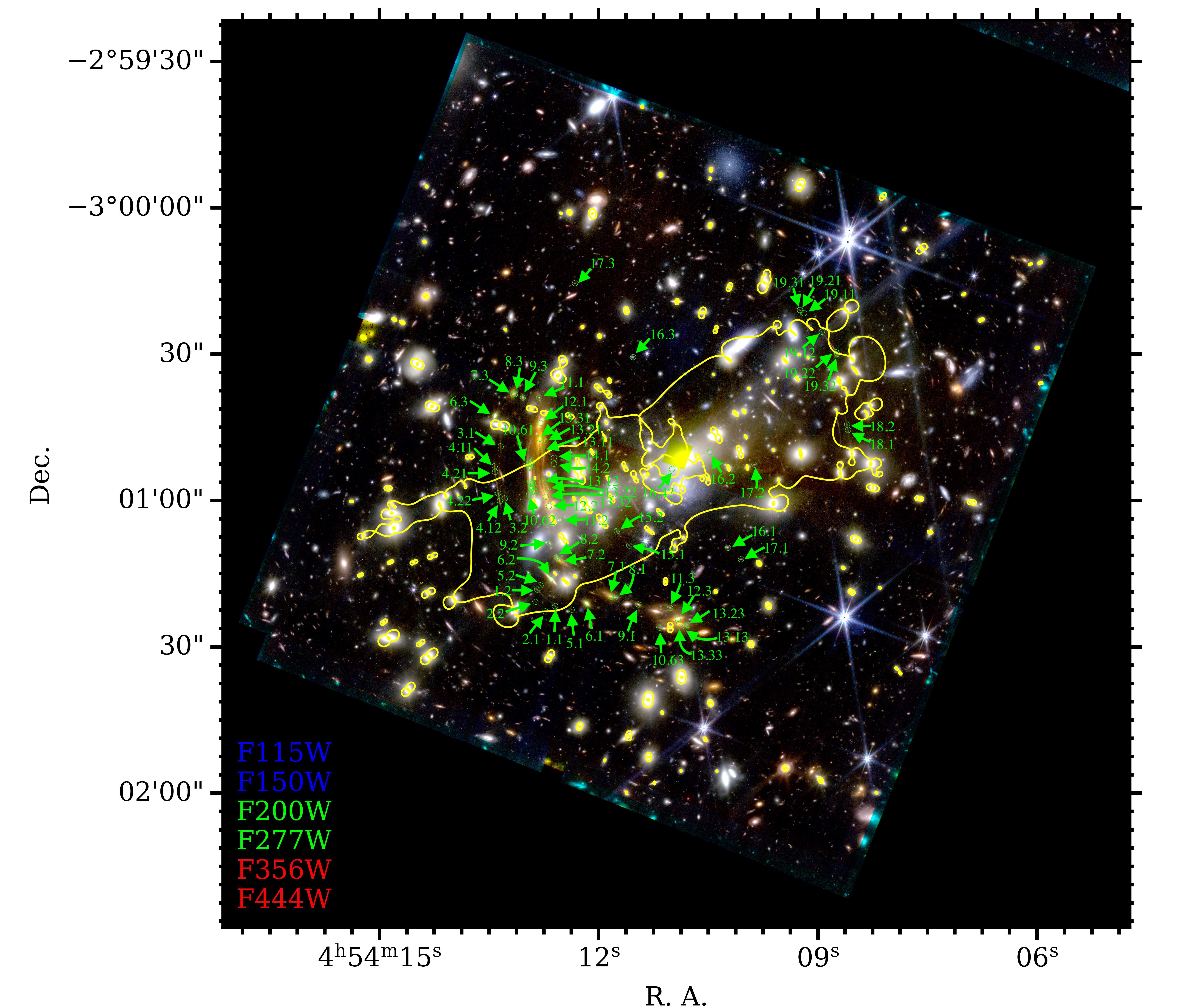}
    \caption{JWST false color image ($R={\rm F356W+F444W}$, $G={\rm F200W+F277W}$, $B={\rm F115W+F150W}$) of MS0451~($z=0.55$) galaxy cluster. The yellow curves show the critical curves for a source at redshift~$z=6.70$, corresponding to \texttt{Zitrin-Analytic} lens model. The lensed images are encircled with thin, dashed green circles, and their positions are further highlighted by green arrows. We did not mark systems~10.1 to~10.5, as they lie very close to each other. The knots in system~19 are also not marked for the same reason. Both systems~10 and~19 correspond to the swallowtail image formations. For a zoomed-in view of these arcs, we refer readers to \Fref{fig:swallow_291} and \Fref{fig:swallow_6p7} below.}
    \label{fig:clus_img}
\end{figure*}

Observing image formations near swallowtail/umbilic caustic structures provides important constraints on the cluster lens mass distribution and allows us to study small-scale structures in the background galaxy sources. For example, image formation near a hyperbolic umbilic provides stringent constraints on the central density profile of galaxy clusters~\citep{2008A&A...489...23L}. At the same time, the large magnification factors allow us to search for subhalos in the vicinity of the observed image formation~\citep{2023MNRAS.522.1091L, 2023MNRAS.526.3902M}. Similarly, due to large magnification factors, swallowtail image formations are excellent targets for observing small-scale details in the background source. We refer readers to Chapter~9 in~\citet{2001stgl.book.....P} for more details on caustic structures around swallowtail and umbilics. Typically, in galaxy-scale lenses, a swallowtail caustic structure appears when a subhalo (which is orders of magnitude smaller in mass compared to main lens) sits close to the critical curve of the main lens~\citep[e.g.,][]{2004A&A...423..797B, 2024SSRv..220...58V}\footnote{It is important to note that this is not a necessary condition. A lens made of two halos with similar masses could also give rise to swallowtail image formations~\citep[\citealt{2015ApJ...806...63D}; also see Fig. 1 in][]{2020MNRAS.492.3294M}.}. However, in such cases, it can be challenging to resolve the swallowtail image formation, and the overall image formation is typically studied as perturbations in the standard, for example, fold image formation with the primary aim being the detection and characterization of subhalos~\citep[e.g.,][]{2025A&A...699A.222D, 2025ApJ...991L..53H}. In galaxy cluster lenses, we can observe swallowtails that lead to the formation of resolved images. In such cases, in addition to enabling subhalo searches, we can also study the small-scale structures in the source galaxy due to large magnification factors. A well-known swallowtail image formation is the Dragon arc~($z=0.7251$) in Abell~370 galaxy cluster~\citep[e.g.,][]{1998MNRAS.294..734A}, which led to a large number of lensed stars in recent JWST observations~\citep{2025NatAs...9..428F}. Another swallowtail image formation at~$z\sim4$ was reported in RX~J1347.5--1145 galaxy cluster~\citep{2014MNRAS.437.1858K} and many more such image formations are expected to be discovered in the near future~\citep{2021MNRAS.503.2097M}.

In this work, we present the discovery of two swallowtail image formations in the~MS~0451.6--0305~($z=0.55$) galaxy cluster detected in the JWST Near Infrared Camera~(JWST-NIRCam) observations. The first swallowtail image formation corresponds to a lensed galaxy at~$z=2.91$, which is part of a submillimetre galaxy group~\citep{2014MNRAS.445..201M}. Only a small region of the source galaxy -- containing at least a pair of knots -- sits inside the swallowtail caustic. We show that the images of these knots reside in very high magnification regions and have extremely small intrinsic source plane sizes~(i.e., effective radii) of~$\lesssim 0.8 - 1.5$~pc. The other swallowtail image formation corresponds to a lensed galaxy at~$z=6.70$~\citep{2016MNRAS.462L...6K, 2021MNRAS.508.1206J}. With JWST, in this arc, we again observe highly magnified compact knots with estimated source plane sizes of~$\lesssim 0.8-18.5$~pc.

The current work is organized as follows. In \Sref{sec:data}, we briefly describe the MS~0451.6--0305 galaxy cluster and its observations. \Sref{sec:modelling} presents an updated parametric strong lensing mass model of MS~0451.6--0305. Our method to measure the sizes of various knots is discussed in \Sref{sec:size}. The swallowtail image formations are discussed in \Sref{sec:swallow_2p91} and \Sref{sec:swallow_6p7}. We have also searched for variability in lensed arcs and discuss it in \Sref{sec:variability}. We conclude this work in \Sref{sec:conclusions}. We use a flat $\Lambda$CDM cosmology with parameters,~$H_0=70\:{\rm km\:s^{-1}\:Mpc^{-1}}$, $\Omega_{m,0}=0.3$, and~$\Omega_\Lambda=0.7$. With the above,~$1''$ is equal to 6.41~kpc at the cluster redshift~($z=0.55$). The age of the Universe at~$z=2.91$ and~$z=6.70$ is 2.18~Gyr and 0.8~Gyr, respectively. All magnitudes are given in AB~system~\citep{1983ApJ...266..713O}.

\begin{figure}
    \centering
    \includegraphics[width=0.45\textwidth]{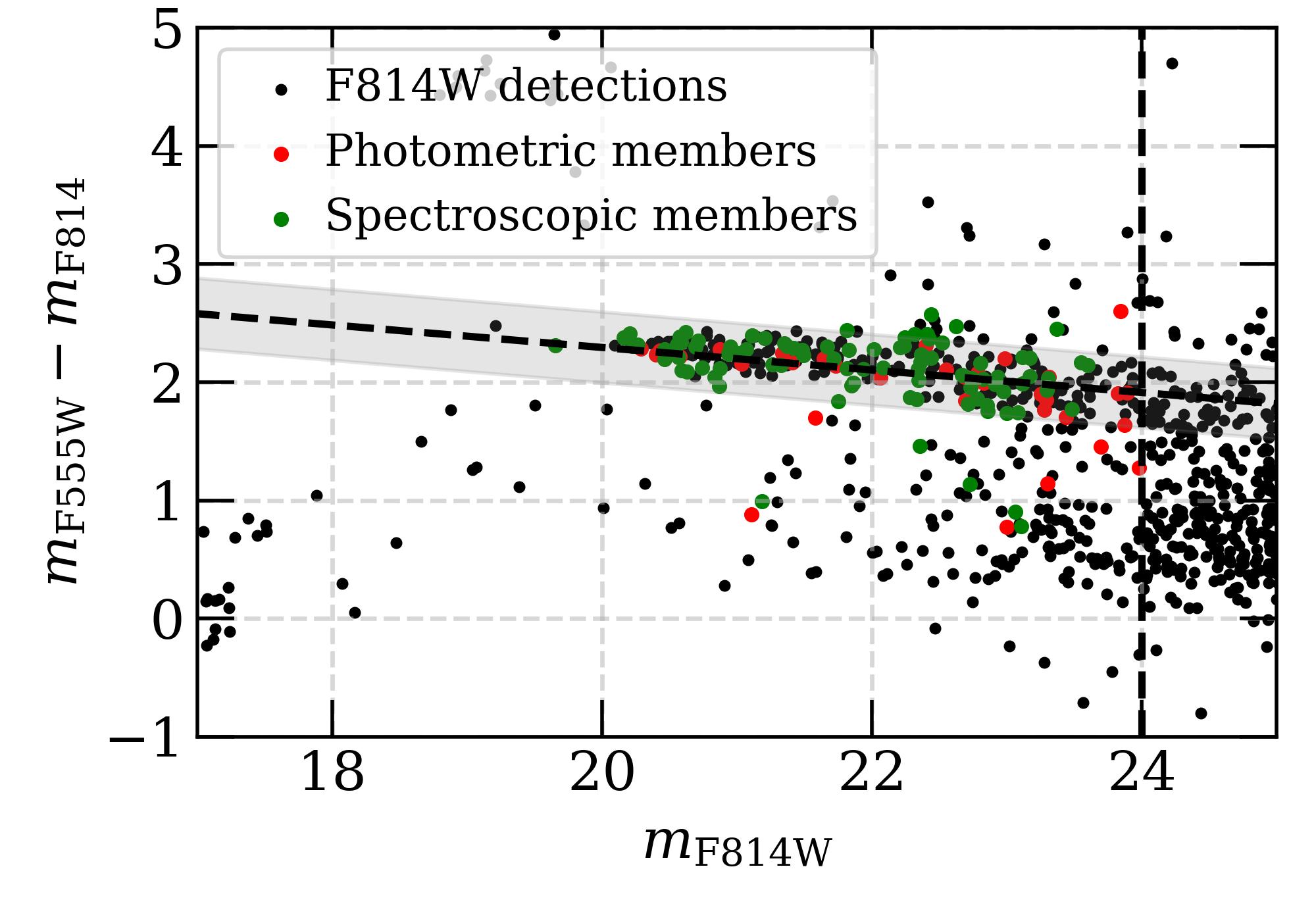}
    \caption{Color-magnitude diagram for objects in MS0451. The black points represent all objects detected in the HST-ACS-F814W filter. The green and red points show the spectroscopic and photometric cluster members, respectively. The cluster red sequence is shown in the gray shaded region. The vertical dashed curve highlights the limiting magnitude for cluster members selected using the red sequence analysis.}
    \label{fig:red_seq}
\end{figure}

\section{MS~0451.6--0305 and JWST data}
\label{sec:data}
MS~0451.6--0305~(MS0451, hereafter) galaxy cluster is an extremely bright X-ray source and was serendipitously discovered by the \textit{Einstein Observatory}~\citep{1991ApJS...76..813S}\footnote{MS0451 was the most luminous X-ray galaxy cluster in the \textit{Einstein Observatory} Extended Medium Sensitivity Survey~(EMSS).}. Later, with optical observations, a giant lensed arc was also detected~\citep{1994ApJS...94..583G, 1999A&AS..136..117L}. The follow-up observations confirmed the nature of the giant arc as a strongly lensed submillimetre galaxy at~$z=2.91$, and further revealed the presence of two additional lensed background galaxies in its vicinity having similar redshifts~\citep[][]{2002MNRAS.330...92C, 2004MNRAS.352..759B}. \citet[][hereafter~\citetalias{2014MNRAS.445..201M}]{2014MNRAS.445..201M} increased the number of identified lensed galaxies at~$z\sim2.91$ to six, identified additional lensed images, and presented an improved lens model for MS0451. 

Recently~\citet[][hereafter~\citetalias{2021MNRAS.508.1206J}]{2021MNRAS.508.1206J}, based on Hubble Space Telescope and Multi-Unit Spectroscopic Explorer~\citep[MUSE;][]{2010SPIE.7735E..08B} observations, presented an updated lensed model based on 16 lensed systems with 46 lensed images. More recently, as part of the JWST cycle-3 GO program~(program ID:~5058, P.I.: L. J. Furtak \& A. K. Meena), MS0451 was observed with JWST-NIRCam in five epochs separated by three to five days between 23/01/2025 to 07/02/2025 UTC. Each epoch targeted MS0451 in six JWST-NIRCam broadband filters~(F115W, F150W, F200W, F277W, F356W, F444W) with an exposure time of 3157~seconds in each filter, reaching $5\sigma$ depths of~$\sim29$~AB. In this work, we utilize the HST and MUSE data products presented in \citetalias{2021MNRAS.508.1206J}, to which the reader is referred for further details. The JWST data were reduced using both the standard JWST pipeline and the \texttt{Grizli} pipeline~\citep{2022zndo...6672538B} for current analyses, and the photometric redshifts for various sources were estimated using \texttt{Eazy}~\citep{2008ApJ...686.1503B}.

\section{Strong lens modeling}
\label{sec:modelling}

\subsection{Cluster galaxies}
\label{ssec:clus_gals}
For the cluster galaxies catalog, following \citetalias{2021MNRAS.508.1206J}, we start by selecting all 112 galaxies with spectroscopic redshifts in the range,~$z\in[0.5307, 0.5652]$ and consider them as part of the galaxy cluster. Next, we complement the above selection with the members selected based on the photometric redshifts. Finally, we perform a color-magnitude analysis using HST-ACS/F555W and HST-ACS/F814W filters, selecting galaxies that lie in the cluster's red sequence~\citep{2000AJ....120.2148G} as shown in \Fref{fig:red_seq}. We fit a linear relation to the colors of spectroscopic members and select galaxies in a color window of~$\Delta m = 0.3$~AB up to 24~AB, which are shown by a dashed curve and a gray shaded region around it, respectively, in \Fref{fig:red_seq}. Ultimately, we conduct a visual inspection to verify the quality of our red-sequence selection. We also include two foreground spiral galaxies when reconstructing the lensing model, as they are located close to the lensed images.

\subsection{Multiple image systems}
\label{ssec:mult_sys}
Before JWST observations of MS0451, with HST observations (along with MUSE), \citetalias{2021MNRAS.508.1206J} identified a total of 16 lensed systems with 46 lensed images. With JWST observations, we have identified eight new lensed systems (one of which is a candidate system). More importantly, we are able to resolve a large number of small-scale structures in pre-existing lensed images. The full list of lensed systems used in our current work is given in \Tref{tab:lensed}, which are also shown in \Fref{fig:clus_img}. It is important to note that we do not utilize all multiple image systems presented in \citetalias{2021MNRAS.508.1206J} as we were unable to unambiguously identify the counter-images. Ultimately, we utilize a total of 19 lensed systems comprising 76 lensed images to reconstruct our lens model(s). We note that the swallowtail image formations correspond to system~10 and system~19.

\begin{figure*}
    \centering
    \includegraphics[width=0.276\textwidth]{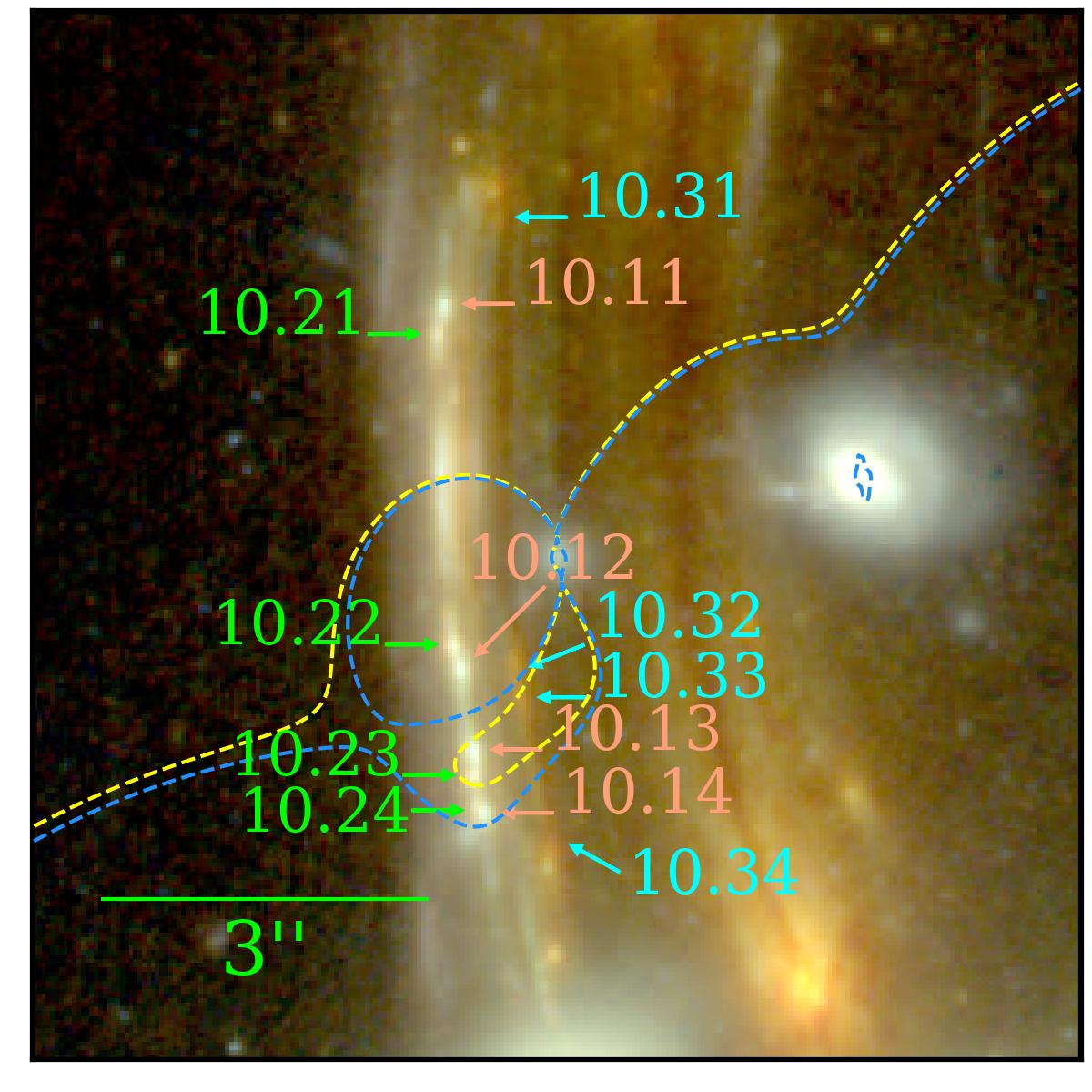}
    \includegraphics[width=0.350\textwidth]{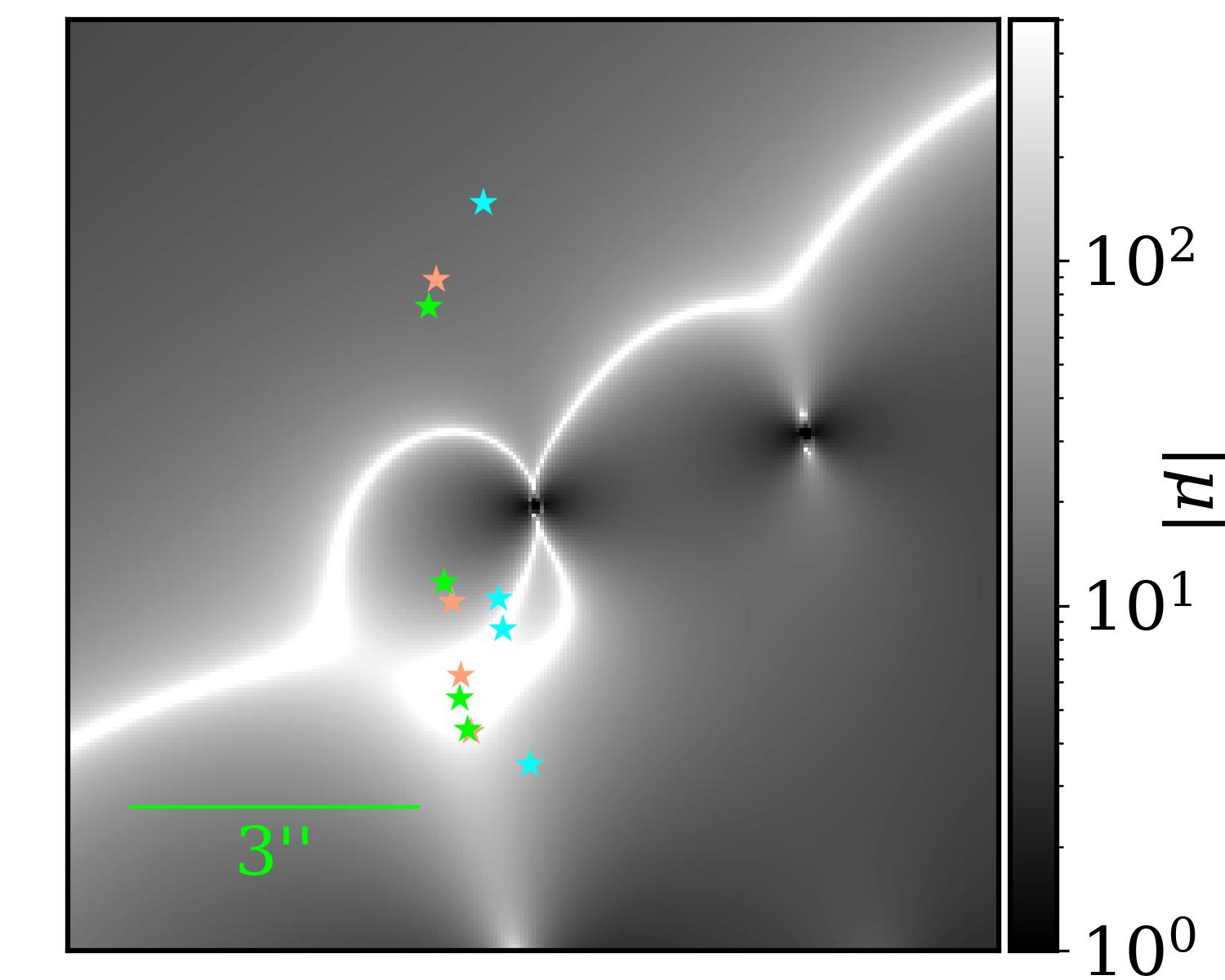}
    \includegraphics[width=0.350\textwidth]{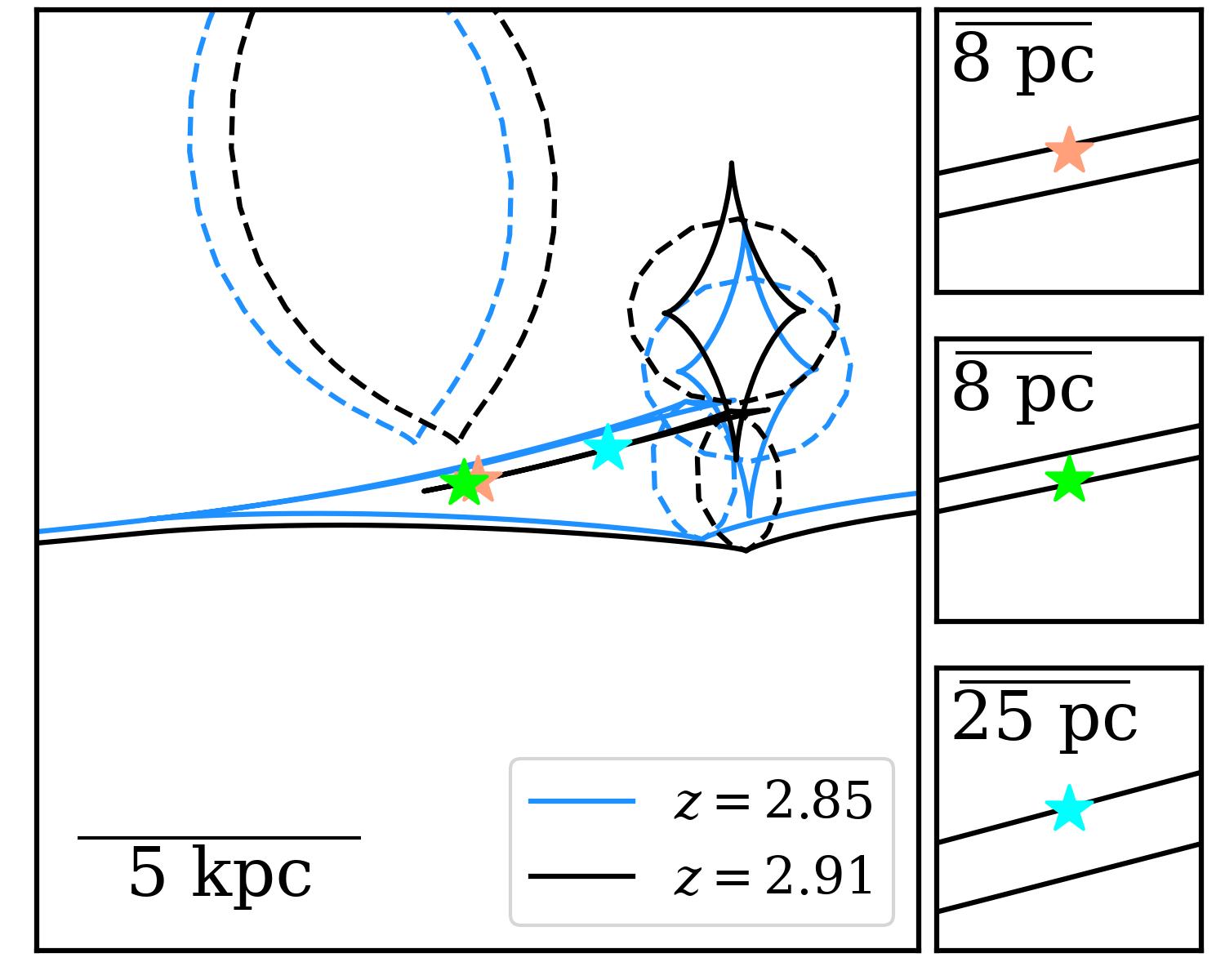}
    \caption{Swallowtail image formation at~$z=2.91$. \textit{Left panel} shows the false color image of the cluster with blue and yellow dashed curves representing the critical curve for~$z=2.85$ and~$z_s=2.91$, respectively. The quadruply imaged knots are marked by colored arrows. \textit{Middle panel} represents the corresponding magnification map with different colored (same as the arrow colors in the left panel) stars marking the observed positions of the quadruply imaged knots. \textit{Right panel} shows the corresponding caustic structure in the source plane. The solid and dashed curves in blue and black show tangential and radial critical caustics at~$z=2.85$ and~$z=2.91$, respectively, showing the evolution of caustics close to the source position(s). The three stars show the barycenter source position for the multiple image systems shown in the left and middle panels. The three inset panels display a zoomed-in view of the regions near each source position. Note that the scale in the right panel shows the proper distance in the source plane.}
    \label{fig:swallow_291}
\end{figure*}

\subsection{Lens model reconstruction}
\label{ssec:sl_model}
We use the parametric strong lensing modelling code \texttt{Zitrin-Analytic}~\citep{2015ApJ...801...44Z, 2023MNRAS.523.4568F} to model MS0451. The underlying mass modeling method has two main components: cluster dark matter halo(s) and cluster galaxies. The dark matter halos are modeled as Pseudo Isothermal Elliptical Mass Distributions~\citep[PIEMD;][]{1983MNRAS.202..995J}, whereas individual cluster galaxies are modeled as dual Pseudo Isothermal Ellipsoid~\citep[dPIE;][]{2007arXiv0710.5636E} with their weights determined by the scaling relations~\citep[e.g.,][]{2011A&ARv..19...47K} with HST-ACS/F814W magnitude as our reference. The foreground spiral galaxies are also assumed to be part of the galaxy cluster, while their weights are left free to vary. The model utilizes Markov Chain Monte Carlo~(MCMC) sampling to optimize the lens model in the source plane such that for a given lensed system, the~$\chi^2$ is defined as the separation between the predicted source positions for each individual image and the corresponding barycenter. The optimization is performed in two steps: first, by running short chains (a few thousand steps) to explore the parameter space, and then running long chains (with $\sim10^5$ steps) for refinement while decreasing the temperature. We refer readers to~\citet{2023MNRAS.523.4568F} for more details about the reconstruction method.

In our current model, we place one dark matter halo at the position of the BCG and another smaller halo at~(73.5616802, -3.0182650), and their center were left free to vary. We run 100 short and 5 long (each with $2\times10^5$ steps) chains to optimize the lens model. Due to the very complex image formations, we also enforce the model to reproduce correct parity for system~10.1, 10.2, 19.1, and 19.2. Although system~10.3 is also identified as a quadruply imaged system (excluding global minimum), we do not enforce parity for this system due to the lack of obvious features in counter-images. For system~19, JWST reveals new small-scale structures in the observed lensed arc. However, due to the complex nature of the underlying image formation, it is challenging to determine the exact multiplicity of different knots. Hence, we only choose lensed knots for which we are confident of counter-images while optimizing the lens model. Our best-fit lens model has image plane root-mean-square~(rms) error of~$0.\arcsec58$. Our model produces various knots in systems 10 and 19 with effectively zero rms\footnote{More precisely, the rms is~$<0.\arcsec04$, i.e., the grid size adapted during lens modeling.}.

\section{Size measurement method}
\label{sec:size}
To determine the sizes of various knots and clumps in lensed arcs, we use full-depth co-added JWST-NIRCam images corresponding to the above-mentioned five epochs of GO program 5058. Because these features are rather small in angular extent, we relied only on NIRCam SW images that were drizzled with a pixel scale of $0.\arcsec03$. We fitted the flux from our targets, nearby sources, and the local background simultaneously, by employing empirical point spread functions~(PSFs) constructed from unsaturated stars in the NIRCam data. The analysis of each target focused on an $11\times 11$ pixel region of interest (ROI), which is large enough to capture the visible area of the source, but small enough to minimize contamination from adjacent sources. For each object, we fitted a two-dimensional elliptical S\'ersic model~\citep{1963BAAA....6...41S}, treating the model parameters~(amplitude, S\'ersic index, effective radius, ellipticity, and position angle), source locations, and local background as free variables. The least squares statistic guided the fit, with optimization achieved through the Trust Region Reflective algorithm \citep{1999SJSC...21....1B}. To estimate the size uncertainties, we combined the variance in fitted S\'ersic effective radii across images from three SW filters with the variance of the size parameter from the fits. For each band, we evaluated the fitting uncertainties by adding the ROI of each source with the local background subtracted into a randomly selected background region within the same image, repeating the fitting process and analyzing the variance from a Monte Carlo sample of these results. The sizes we measured correspond to the apparent effective angular radii of the knots.

For many of the knots, the effective radius of the S\'ersic profile falls below a certain threshold. In such cases, the PSF-convolved image becomes indistinguishable from the PSF itself. In such cases, the method loses its sensitivity to the size of a source, and it effectively becomes a point source. To determine the detection limit for point sources, we employed the method on a collection of unsaturated stars within the cluster field. We calculated the detection limit for point sources by analyzing the 68\% containment of the sizes measured from these effective point sources. In the results for the knots and clumps in the arc, if the lower bounds of their sizes are smaller than the point-source limit, we only provide upper bounds.

\begin{deluxetable}{ccccc}[!ht]
\label{tab:z291}
\tablecaption{Size measurements for knots at $z=2.91$.}
\tablewidth{\columnwidth}
\tablehead{
\colhead{ID}  & \colhead{$[\mu_t, \mu_r]$} & \colhead{$\theta_{\rm image}$} & \colhead{$\theta_{\rm source}$} & \colhead{$R_{\rm source}$} \\
              &                            & \colhead{(in mas)}       & \colhead{(in mas)}        & \colhead{(in pc)} }
\startdata
10.11  & [10.0, 1.5]   & $97.9 \pm 35.1$  & $9.8 \pm 3.5$ & $75.9 \pm 27.2$ \\
10.12  & [-37.2, 1.6]  & $154.4 \pm 25.0$ & $4.2 \pm 0.7$ & $32.3 \pm 5.2$  \\
10.13  & [1355.7, 1.7] & $116.4 \pm 47.3$ & $0.1 \pm 0.0$ & $0.7 \pm 0.3$   \\
10.14  & [-329.4, 1.8] & $45.0 \pm 12.3$  & $0.1 \pm 0.0$ & $1.1 \pm 0.3$   \\
10.21  & [11.0, 1.5]   & $<33.9$          & $<3.1$        & $<23.9$         \\
10.22  & [-22.9, 1.5]  & $<35.5$          & $<1.6$        & $<12.0$         \\
10.23  & [1764.6, 1.7] & $<36.9$          & $<0.02$       & $<0.2$          \\
10.24  & [-386.2, 1.8] & $<39.7$          & $<0.1$        & $<0.8$          \\
\enddata
\tablecomments{Column~(1): Knot system ID. Column~(2): Best-fit tangential and radial magnifications at the observed lensed image positions. Column~(3): $\theta_{\rm image}$ represents the measured/observed effective angular radius of the knot in milli-arcseconds. Column~(4): $\theta_{\rm source}$ represents the lensing corrected (i.e., observed size divided by tangential magnification) source plane effective angular radius of the knot in milli-arcseconds. Column~(5): $R_{\rm source}$ represents the lensing corrected source plane effective radius of the knot in parsecs. For a PSF-dominated source, we only provide upper bounds.}
\end{deluxetable}

\section{Swallowtail image formation at~$\lowercase{z}=2.91$}
\label{sec:swallow_2p91}
MS0451 strongly lenses a submillimetre galaxy group at $ z\simeq2.91$. This group consists of at least seven galaxies, out of which six are strongly lensed and are represented by system-6 to system-13 in \Tref{tab:lensed}. This galaxy group is reported to have a large star formation rate of~$450\pm50~{\rm M_\odot/yr}$, which is primarily driven by the galaxies undergoing a merger~(see \citetalias{2014MNRAS.445..201M} for more details).

\begin{figure*}[!ht]
    \centering
    \includegraphics[width=0.276\textwidth]{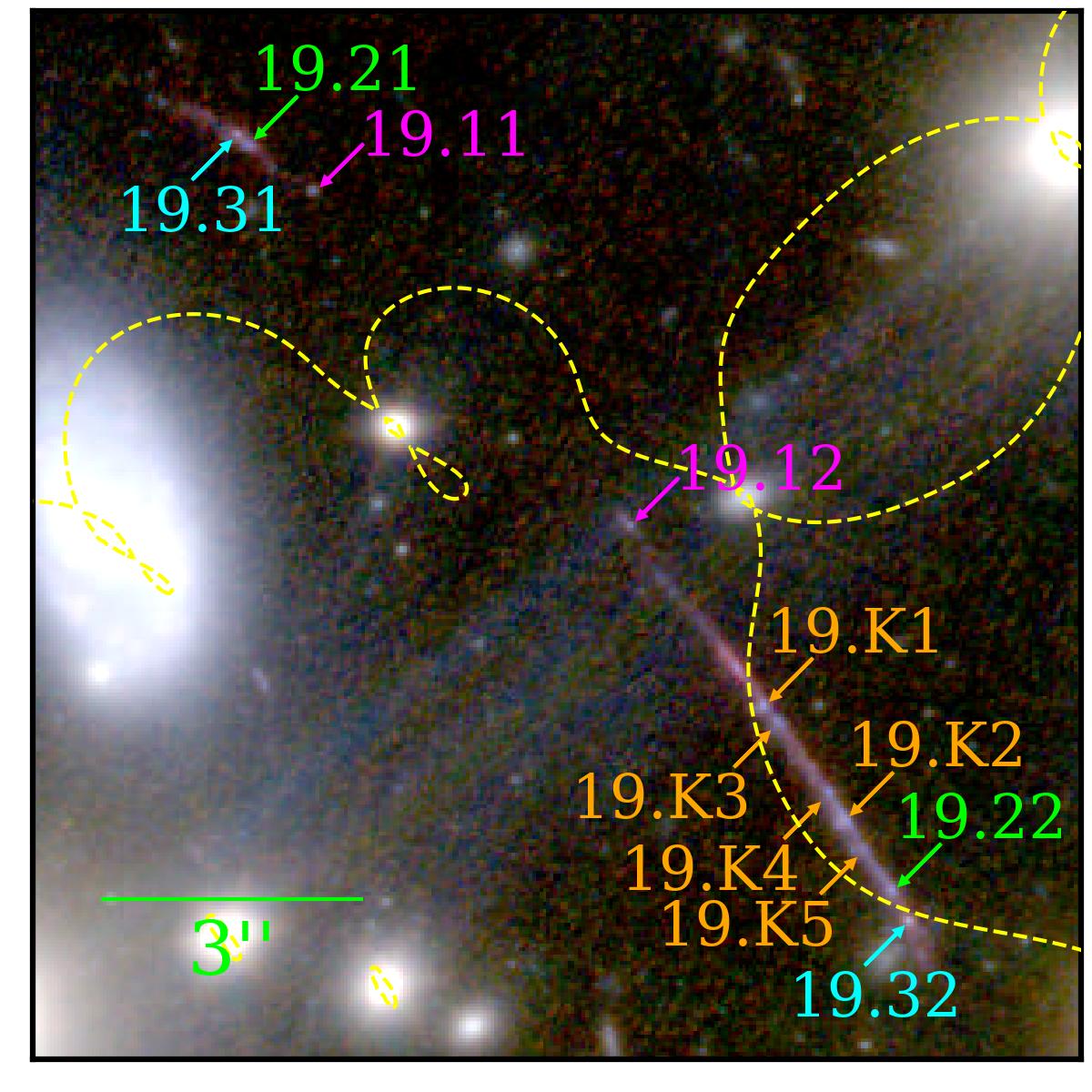}
    \includegraphics[width=0.350\textwidth]{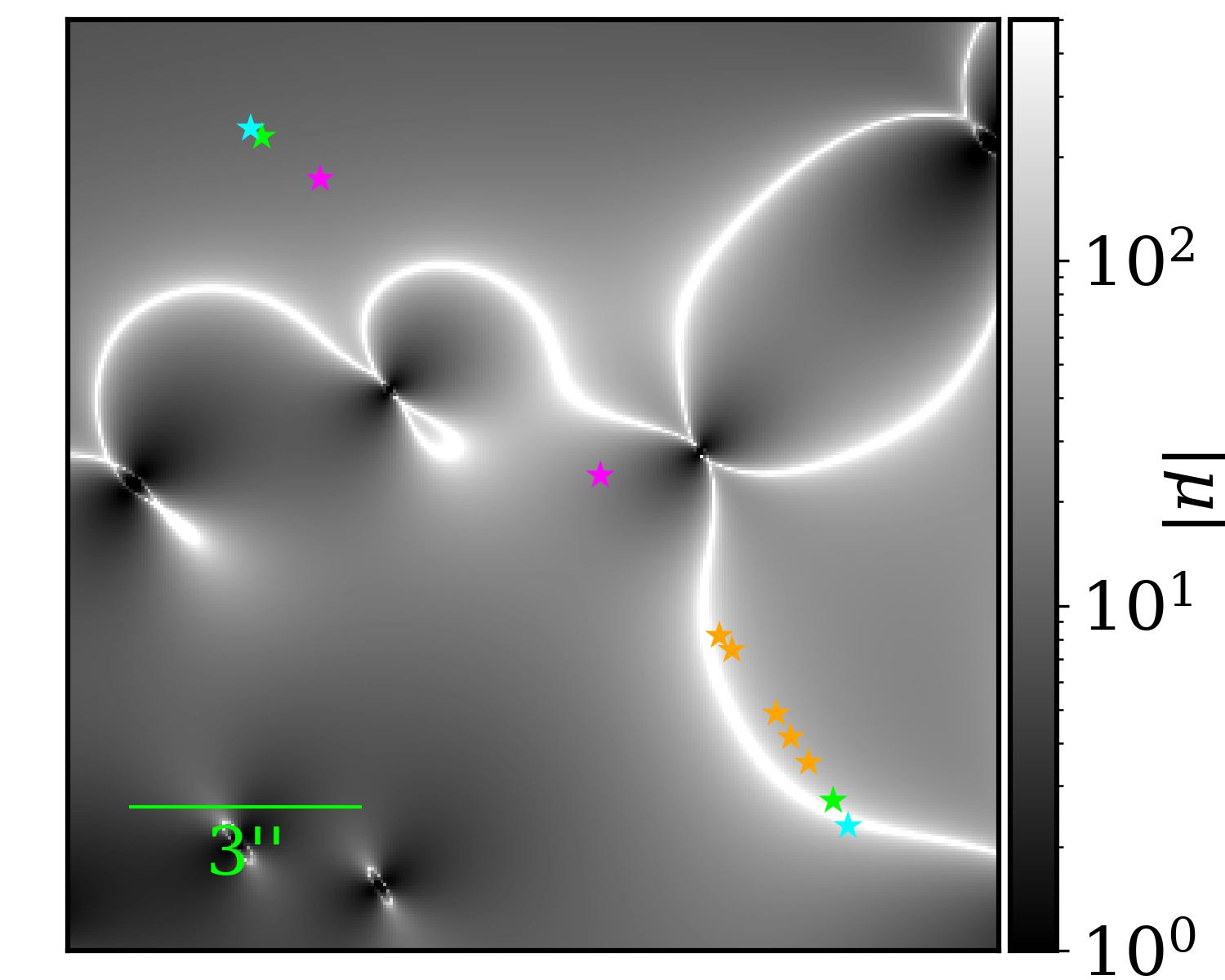}
    \includegraphics[width=0.350\textwidth]{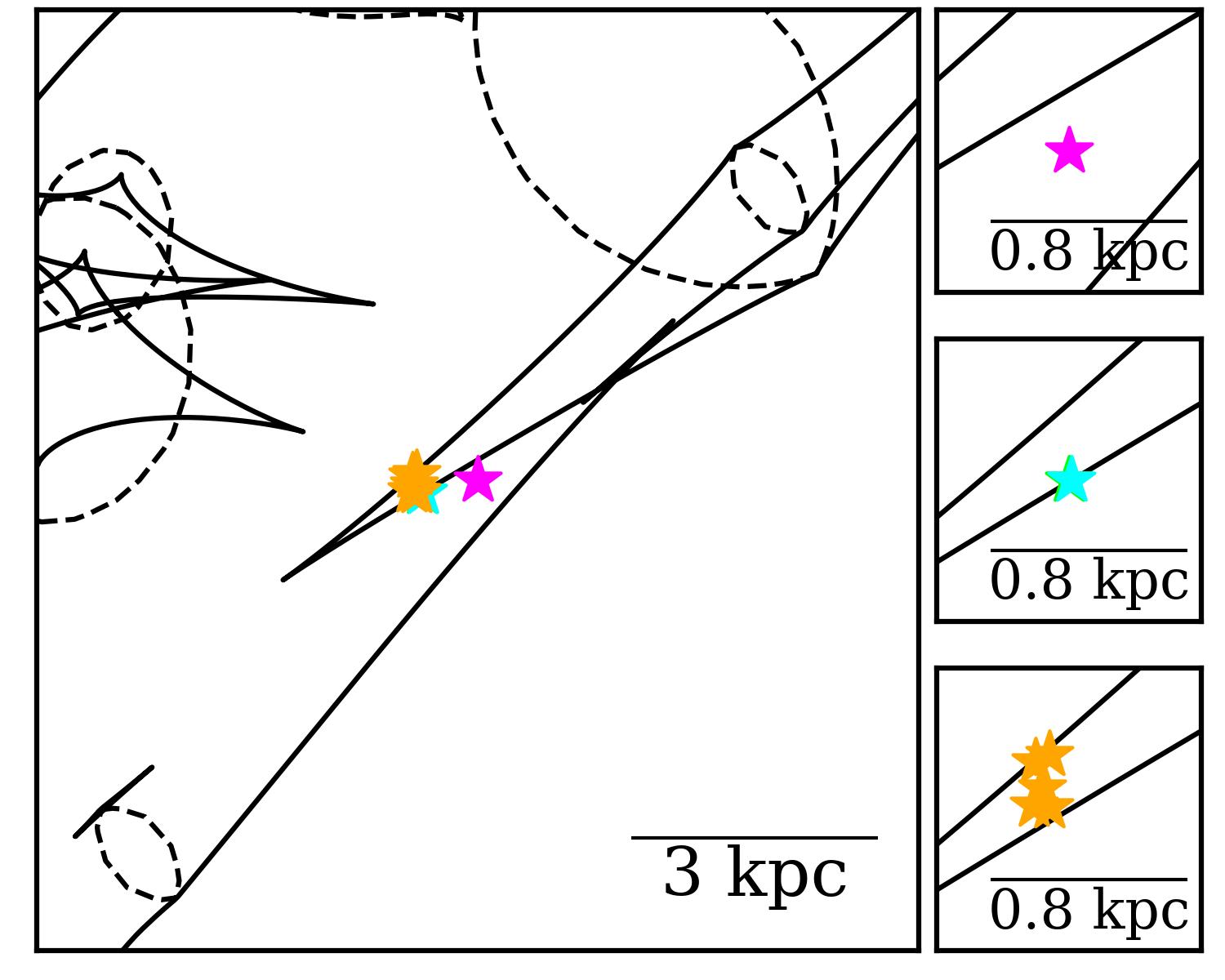}
    \caption{Swallowtail image formation at~$z=6.70$. \textit{Left panel} shows the false color image of the cluster with yellow dashed curves representing the critical curve for~$z_s=6.70$. Different systems and knots are shown in different colors. We note that 19.K1/K2/K3/K4/K5 are not counter-images of each other; instead, they are individual knots. See \Sref{sec:swallow_6p7} for more details. \textit{Middle panel} shows the corresponding magnification map with different colored (same color scheme as the left panel) stars representing the observed positions of the different knots. \textit{Right panel} shows the caustic structure in the source plane with solid and dashed curves representing the tangential and radial caustics, respectively. The stars indicate the barycenter position of different multiple-image systems shown in the left panel. Note that the scale in the right panel shows the proper distance in the source plane.}
    \label{fig:swallow_6p7}
\end{figure*}

A small part of one of these galaxies (system 10) lies within the swallowtail caustic, resulting in the formation of an arc composed of four images. As we can see from~\Fref{fig:swallow_291}, at least two knots are quadruply imaged, which are marked as system 10.1 and 10.2. We also identified a third knot, corresponding to system 10.3, which may also be sitting inside the swallowtail caustic (as also predicted by our best-fit lens model). That said, we did not force the model to reproduce its parity, as the corresponding swallowtail image formation is not as clear as the other two knots. From the left panel in \Fref{fig:swallow_291}, we can see that 10.13/23 and 10.14/24 are very highly magnified as they sit very close to the caustic (see \Tref{tab:z291} for point source magnification values). The high magnification for 10.13/23 and 10.14/24 knots can also be inferred from the fact that they sit very close to each other, and the critical curve has to pass between them. Even then, the pair is well resolved, implying the very compact nature of these knots. 

In the right panel of~\Fref{fig:swallow_291}, we show the caustics in the source plane for~$z=2.85$ and~$z=2.91$ in blue and black, respectively. We show caustics for two source redshifts to present the evolution of swallowtail caustic structure. We can see that at~$z=2.85$, close to the predicted source positions (indicated by stars), the caustics show the characteristic swallowtail structure. As we increase source redshift to~$z=2.91$, part of the swallowtail caustic gets detached from the main caustic spine. We also note that this detached caustic component spans only a very narrow region along the ordinate axis, as also seen in the inset plots. Since we can spatially distinguish the different lensed counterparts of these knots, we can infer that these knots must lie entirely within the narrow caustic structure; if any part of a knot crossed the caustic, it would produce overlapping images. Hence, from \Fref{fig:swallow_291} alone, we can infer that the source plane sizes of knots corresponding to system~10.1 and~10.2 should be~$\lesssim 3 - 4$~pc. We obtain more quantitative constraints on source effective radii by fitting S\'ersic profiles to all of the counter-images of these knots (see~\Sref{sec:size} for more details) and are listed in \Tref{tab:z291}. We note that counter-images of system 10.2 have magnification values ranging from~$10-2000$ and even for the image with the most magnification, 10.23, the observed effective radius is PSF-dominated, again hinting at a highly compact intrinsic nature of the source. As expected, the lensing-corrected source plane size estimates vary based on the corresponding tangential magnification values, with the most stringent constraints on the size coming from the most magnified knots. Hence, from \Tref{tab:z291}, we can see that both systems 10.1 and 10.2 correspond to knots with source plane effective radii of~$\lesssim1.0$~pc. It is important to note that at higher magnification values, the associated errors would also be large. Therefore, for a more conservative estimate, even if we consider the second most magnified counter-images from systems 10.1 and 10.2, the above inferences that these sources have (sub-)parsec sizes remain valid.

\begin{deluxetable}{ccccc}[!ht]
\label{tab:z6p70}
\tablecaption{Size measurements for knots at $z=6.70$.}
\tablewidth{\columnwidth}
\tablehead{
\colhead{ID}  & \colhead{$[\mu_t, \mu_r]$} & \colhead{$\theta_{\rm image}$} & \colhead{$\theta_{\rm source}$} & \colhead{$R_{\rm source}$} \\
              &                            & \colhead{(in mas)}       & \colhead{(in mas)}        & \colhead{(in pc)} }
\startdata
19.12  & [-19.6, 1.2]  & $<40.3$          & $<2.1$        & $<11.0$        \\
19.22  & [120.5, 1.5]  & $<30.6$          & $<0.3$        & $<1.3$         \\
19.32  & [-211.9, 1.6] & $<32.7$          & $<0.2$        & $<0.8$         \\
19.K1  & [149.9, 1.3]  & $<30.2$          & $<0.2$        & $<1.1$         \\
19.K2  & [52.8, 1.4]   & $129.4 \pm 53.0$ & $2.4 \pm 1.0$ & $13.1 \pm 5.4$ \\
19.K3  & [90.4, 1.4]   & $<32.7$          & $<0.4$        & $<1.9$         \\
19.K4  & [52.7, 1.4]   & $<36.3$          & $<0.7$        & $<3.7$          \\
19.K5  & [56.8, 1.5]   & $78.9 \pm 24.4$  & $1.4 \pm 0.4$ & $7.4 \pm 2.3$  \\
\enddata
\tablecomments{Column~(1): Knot system ID. Column~(2): Best-fit tangential and radial magnifications at the observed lensed image positions. Column~(3): $\theta_{\rm image}$ represents the measured/observed effective angular radius of the knot in milli-arcseconds. Column~(4): $\theta_{\rm source}$ represents the lensing corrected (i.e., observed size divided by tangential magnification) source plane effective angular radius of the knot in milli-arcseconds. Column~(5): $R_{\rm source}$ represents the lensing corrected source plane effective radius of the knot in parsecs. For a PSF-dominated source, we only provide upper bounds.}
\end{deluxetable}

It is important to emphasize the role of swallowtail caustic (and JWST imaging) in the above constraints on the size measurements. Earlier models, presented in \citetalias{2014MNRAS.445..201M} and \citetalias{2021MNRAS.508.1206J}, considered the lensed arc as a double image formation. In such cases, we observe that the critical curve only crosses the arc once. However, with JWST imaging, we are able to resolve the underlying image formation and determine that the critical curve crosses the overall arc three times, and a large part of the arc is highly magnified. In addition, with the extremely narrow width of the swallowtail caustic alone, we can get much tighter constraints on the source size compared to 10.11/12 and 10.21/10.22 counter-images. Size constraints coming from 10.11/12 and 10.21/22 are also representative of the case in which the arc would instead correspond to a simple double-image formation, highlighting the importance of the swallowtail caustic in current size estimations. At the same time, it is also important to note that not every swallowtail caustic structure necessarily provides such tight constraints, as we will see in the following section.

\section{Swallowtail image formation at~$\lowercase{z}=6.70$}
\label{sec:swallow_6p7}
In earlier work, \citetalias{2021MNRAS.508.1206J} highlighted the quadruply imaged nature (excluding global minimum) of the~$z=6.70$ arc, although the authors did not go into details of it being a swallowtail image formation. With JWST imaging, we can resolve multiple small-scale knots in this arc; however, identifying different counter-images remains challenging. Therefore, in our lens model reconstruction, we only use three image pairs~(system 19.1, 19.2, 19.3) for which the counter-images can be confidently identified, and are highlighted in the left panel of \Fref{fig:swallow_6p7}.

The newly identified knots in this arc can also be found in \Fref{fig:swallow_6p7}. We find that our best-fit lens model, similar to that of~\citetalias{2021MNRAS.508.1206J}, leads to the swallowtail image formation, which can also be seen from the caustic structure in the right panel of \Fref{fig:swallow_6p7}. Since the corresponding critical curve runs somewhat parallel to the observed arc, the estimated magnification values for different knots would be large. That being said, due to the large size of swallowtail caustics, unlike the previous case, lensing alone cannot provide stringent constraints on their sizes as can be seen from the right panel in \Fref{fig:swallow_6p7}. The effective radii of various knots obtained with the S\'ersic profile fitting are given in \Tref{tab:z6p70}. We can see that many of the knots in this lensed arc are PSF-dominated; hence, we only provide upper bounds on their effective radii. The best-fit model magnification for different knots on the arc varies in the~$[20, 200]$ range, and after correcting for lensing magnification, the source plane knot~(physical) effective radii range between~$\sim0.8-18.5$~pc.

Owing to the complexity of the image formation, it is difficult to determine whether some of the individual knots (labeled as 19.K) are counter-images of one another. If this is indeed the case, it would shift the position of the critical curves relative to the arc, as well as the locations where the critical curve intersects the arc, ultimately modifying the size estimates. One possible way to overcome this is to construct multiple lens models assuming different multiplicities for various knots. We constructed two additional lens models assuming that, (i) model-1: instead of 19.22, 19.K2 is the counter-image of 19.21, (ii) model-2: 19.K1~(19.K3) is a counter-image of 19.K2~(19.K4) and force the parity of these new systems. For model-1, the overall rms was larger to~$0.\arcsec68$ but for model-2, the overall fit improved with rms of~$0.\arcsec37$. The corresponding size estimates are given in \Tref{tab:z6p70_app}. We note that both model-1 and model-2 show decrease in the estimated effective radii for various knots, with this decrease being more pronounced in model-2.

\section{Variability and search for transients}
\label{sec:variability}
The primary goal of the multi-epoch (five times within a month) JWST imaging of MS0451 was to search for transient stars in these high-redshift lensed arcs and measure their properties~\citep{2024jwst.prop.5058F}. In this paper, we specifically search for transients in these two arcs at $z=2.91$ and~$z=6.70$ and examine the variability of the knots and blobs within the arcs. A detailed search for transients in the entire cluster field is reserved for another paper.

To search for transients, we employ the image differencing method. We compare the difference images between each pair of adjacent epochs of images and between each single epoch exposure and the coaddition of the full depth. Although we observe flux fluctuations within these lensed arcs throughout the five epochs of observation, we have not found any prominent transients with a point-source detection significance greater than $2\sigma$. In addition, we also examine the recent JWST-NIRCam observations of MS0451 taken in September~2025 as part of the Vast Exploration for Nascent, Unexplored Sources (VENUS) program~\citep[GO 6882, P.I.: Seiji Fujimoto;][]{2025jwst.prop.6882F} and generate the difference images between the images from GO-5058 and GO-6882. Again, we do not find any significant transients within these lensed arcs. It is important to note that the GO-6882 observations of MS0451 are shallower~\citep[$m_{\rm AB}\sim 28$;][]{2025jwst.prop.6882F} than the GO-5058 program, which also limits us from finding long-term faint transients.

\section{Conclusions}
\label{sec:conclusions}
In this work, we have presented the discovery of two swallowtail image formations in JWST-NIRCam imaging of MS0451 obtained under the JWST-Cy3 GO program (program ID: 5058, P.I.: L. J. Furtak \& A. K. Meena). This is the first instance of observing two swallowtail image formations behind a single galaxy cluster. One of these image formations corresponds to a galaxy that is part of a submillimeter galaxy group at $ z=2.91$, whereas the other swallowtail image formation corresponds to a galaxy at~$z=6.70$. In both of these image formations, we detect individual knots with point source magnifications in~$\sim[10^2,10^3]$ range. These large magnification factors, together with their PSF-dominated observed sizes, indicate that the knots are extremely compact, probing physical scales down to sub-parsec scales in the source plane. As expected for swallowtail caustic, the critical curves show multiple crossings on both of these arcs and lead to high magnification factors for large parts of these sources. Although we did not detect lensed stars or any other lensed transients in these systems, the large magnification factor~(along with the large star formation rate in~$z=2.91$ system; \citetalias{2014MNRAS.445..201M}) makes the MS0451 galaxy cluster a compelling target for future lensed transients searches. 

Motivated by their rarity, this work primarily focuses on the discovery of swallowtail image formations and their potential to probe small-scale structures in the source. As discussed above, high magnification factors associated with swallowtail image formations also imply that the associated errors would also be large. Consequently, small shifts in the position and orientation of the critical curve relative to the arc can lead to considerable variation in the inferred magnification at the observed knot positions. Hence, a detailed investigation of the source properties of these knots would require multiple lens models (preferably both parameteric and non-parametric) to sample the systematic and statistical errors in magnification. These multiple-lens models, along with the detailed properties of lensed galaxies (and various knots), will also allow us to study the prospects of lensed transients in this cluster and will be presented in a follow-up paper.

\section{Acknowledgments}
AKM acknowledges the support from the Start-up Grant IE/CARE-25-0305 provided by the IISc, Bengaluru, India. AZ acknowledges support by the Israel Science Foundation Grant No. 864/23. MJ acknowledges support by the United Kingdom Research and Innovation (UKRI) Future Leaders Fellowship `Using Cosmic Beasts to uncover the Nature of Dark Matter' (grant number MR/X006069/1). PLK acknowledges funding from STScI GO-5058 and NSF AAG 2308051. RAW acknowledges support from NASA JWST Interdisciplinary Scientist grants NAG5-12460, NNX14AN10G, and 80NSSC18K0200 from GSFC. This research was supported by the International Space Science Institute (ISSI) in Bern, through ISSI International Team project \#476 (Cluster Physics From Space To Reveal Dark Matter). This research has made use of NASA’s Astrophysics Data System~(ADS) Bibliographic Services. 

\noindent
\textit{Facilities:} JWST~(NIRCam) and HST~(ACS, WFC3).

\noindent
\textit{software:} \texttt{Python}~(\url{https://www.python.org/}), \texttt{NumPy}~\citep{2020Natur.585..357H}, \texttt{Astropy}~\citep{2022ApJ...935..167A}, \texttt{Matplotlib}~\citep{2007CSE.....9...90H}, \texttt{Grizli}~\citep{2022zndo...6672538B}, \texttt{Eazy}~\citep{2008ApJ...686.1503B}

\startlongtable
\begin{deluxetable}{ccccc}
\label{tab:lensed}
\tablecaption{Multiple image systems in MS0451.}
\tablewidth{\columnwidth}
\tablehead{
\colhead{ID}  & \colhead{R. A.} & \colhead{Dec.} & \colhead{$z_{\rm phot}$} & \colhead{Comments} \\
        & \colhead{deg/J2000} & \colhead{deg/J2000} &               &   }
\startdata
1.1    & 73.5524827   & -3.0226995   & $2.82^{+0.33}_{-0.15}$  & New system \\
1.2    & 73.5535107   & -3.0217571   & $2.85^{+0.24}_{-2.55}$  &            \\
\hline
2.1    & 73.5530417   & -3.0230025   & --                      & New system \\
2.2    & 73.5536163   & -3.0224577   & --                      &            \\
\hline
3.1    & 73.5555895   & -3.0135955   & $4.31^{+0.42}_{-3.06}$  & New system \\
3.2    & 73.5553568   & -3.0165605   & $15.81^{+3.43}_{-0.99}$ &            \\
\hline
4.11   & 73.5559301   & -3.0147271   & $11.72^{+0.25}_{-0.29}$ & New system \\
4.12   & 73.5556421   & -3.0167149   & $3.28^{+0.18}_{-0.17}$  &            \\
\cline{1-1}
4.21   & 73.5559138   & -3.0150589   & $2.75^{+0.19}_{-0.25}$  &            \\
4.22   & 73.5557350   & -3.0162756   & $0.50^{+0.31}_{-0.09}$  &            \\
\hline
5.1    & 73.5515458   & -3.0228975   & --                      & New system \\
5.1    & 73.553315    & -3.0215139   & --                      &            \\
\hline
6.1    & 73.55068     & -3.0225506   & 2.93$^\star$            & \citetalias{2014MNRAS.445..201M}, \citetalias{2021MNRAS.508.1206J} \\
6.2    & 73.5527027   & -3.0211649   &                         &            \\
6.3    & 73.555901    & -3.0118641   &                         &            \\
\hline
7.1    & 73.5493191   & -3.0221666   & $2.97^{+0.13}_{-0.13}$  & \citetalias{2014MNRAS.445..201M}, \citetalias{2021MNRAS.508.1206J} \\
7.2    & 73.5522767   & -3.0200823   & $2.74^{+0.12}_{-0.14}$  &            \\
7.3    & 73.5548849   & -3.0106373   & $2.81^{+0.23}_{-0.20}$  &            \\
\hline
8.1    & 73.5490839   & -3.0222040   & --                      & \citetalias{2014MNRAS.445..201M}, \citetalias{2021MNRAS.508.1206J} \\
8.2    & 73.5523868   & -3.0199143   & --                      &            \\
8.3    & 73.5547783   & -3.0106257   & --                      &            \\
\hline
9.1    & 73.5477412   & -3.0226070   & $2.86^{+0.12}_{-0.31}$  & \citetalias{2014MNRAS.445..201M}, \citetalias{2021MNRAS.508.1206J} \\
9.2    & 73.5527843   & -3.0191314   & $1.80^{+0.26}_{-0.65}$  &            \\
9.3    & 73.5543215   & -3.0108163   & $2.74^{+0.11}_{-0.11}$  &            \\
\hline
10.11  & 73.5539325   & -3.0147472   & 2.91$^\star$            & \citetalias{2014MNRAS.445..201M}, \citetalias{2021MNRAS.508.1206J} \\
10.12  & 73.5538875   & -3.0156694   &                         &            \\
10.13  & 73.553862    & -3.0158811   &                         &            \\
10.14  & 73.553833    & -3.0160421   &                         &            \\
\cline{1-1}
10.21  & 73.5539541   & -3.0148243   &                         &            \\
10.22  & 73.5539099   & -3.0156142   &                         &            \\
10.23  & 73.5538651   & -3.0159464   &                         &            \\
10.24  & 73.5538424   & -3.0160359   &                         &            \\
\cline{1-1}
10.31  & 73.5537974   & -3.0145272   &                         &            \\
10.32  & 73.5537542   & -3.0156606   &                         &            \\
10.33  & 73.5537412   & -3.0157483   &                         &            \\
10.34  & 73.5536638   & -3.0161361   &                         &            \\
\cline{1-1}
10.41  & 73.5539308   & -3.0145850   &                         &            \\
10.42  & 73.5538096   & -3.0162067   &                         &            \\
\cline{1-1}
10.51  & 73.5538093   & -3.0144526   &                         &            \\
10.52  & 73.5536763   & -3.0161700   &                         &            \\
\cline{1-1}
10.61  & 73.5540780   & -3.0146280   &                         &            \\
10.62  & 73.5539776   & -3.0163534   &                         &            \\
10.63  & 73.5465506   & -3.0239352   &                         &            \\
\hline
11.1   & 73.5533981   & -3.0108247   & $2.78^{+0.12}_{-0.09}$  & \citetalias{2014MNRAS.445..201M}, \citetalias{2021MNRAS.508.1206J} \\
11.2   & 73.5521849   & -3.0177928   & $2.33^{+0.04}_{-0.23}$  &            \\
11.3   & 73.5459287   & -3.0228025   & $2.66^{+0.20}_{-0.22}$  &            \\
\hline
12.1   & 73.5533304   & -3.0122696   & $2.76^{+0.12}_{-0.12}$  & \citetalias{2014MNRAS.445..201M}, \citetalias{2021MNRAS.508.1206J} \\
12.2   & 73.5528371   & -3.0170077   & $2.91^{+0.13}_{-0.11}$  &            \\
12.3   & 73.5455010   & -3.0234268   & $2.82^{+0.12}_{-0.13}$  &            \\
\hline
13.11  & 73.5532182   & -3.0137917   & --                      & \citetalias{2014MNRAS.445..201M}, \citetalias{2021MNRAS.508.1206J} \\
13.12  & 73.5531578   & -3.0155610   & --                      &            \\
13.13  & 73.5450718   & -3.0238566   & --                      &            \\
\cline{1-1}
13.21  & 73.5530968   & -3.0133016   & --                      &            \\
13.22  & 73.5528940   & -3.0159917   & --                      &            \\
13.23  & 73.5449517   & -3.0237591   & --                      &            \\
\cline{1-1}
13.31  & 73.5533805   & -3.0131750   & $2.44^{+1.05}_{-0.11}$  &            \\
13.32  & 73.5530201   & -3.0164695   & $2.88^{+0.17}_{-0.12}$  &            \\
13.33  & 73.5454140   & -3.0237047   & $3.05^{+0.30}_{-0.25}$  &            \\
\hline
14.1   & 73.5525629   & -3.0142642   & --                      & New system \\
14.2   & 73.5525491   & -3.0145939   & --                      &            \\
\hline
15.1   & 73.5482618   & -3.0192528   & $2.18^{+0.11}_{-0.14}$  & \citetalias{2021MNRAS.508.1206J} \\
15.2   & 73.5489762   & -3.0184306   & --                      &            \\
\hline
16.1   & 73.5426242   & -3.0193765   & $1.75^{+0.11}_{-0.1}$   & \citetalias{2021MNRAS.508.1206J} \\
16.2   & 73.5436542   & -3.0139172   & --                      &            \\
16.3   & 73.5480344   & -3.0085242   & $1.85^{+0.16}_{-0.16}$  &            \\
16.4$\dagger$  & 73.5458233   & -3.0149022   & --              &            \\
\hline
17.1   & 73.5418828   & -3.0200351   & $6.18^{+0.15}_{-0.14}$  & \citetalias{2021MNRAS.508.1206J} \\
17.2   & 73.5411287   & -3.0146485   & $6.86^{+0.07}_{-0.08}$  &            \\
17.3   & 73.5513510   & -3.0043125   & $6.86^{+0.11}_{-0.12}$  &            \\
\hline
18.1   & 73.5357823   & -3.0126723   & $4.86^{+0.26}_{-0.42}$  & New system \\
18.2   & 73.5358352   & -3.0123625   & $4.94^{+0.22}_{-0.49}$  &            \\
\hline
19.11  & 73.5382847   & -3.0060172   & 6.70$^\star$            & \citetalias{2021MNRAS.508.1206J} \\
19.12  & 73.5372804   & -3.0070792   &                         &            \\
\cline{1-1}
19.21  & 73.5384940   & -3.0058659   &                         &            \\
19.22  & 73.5364455   & -3.0082430   &                         &            \\
\cline{1-1}
19.31  & 73.5385341   & -3.0058364   &                         &            \\
19.32  & 73.5363924   & -3.0083343   &                         &            \\
\cline{1-1}
19.K1  & 73.5368537   & -3.0076528   &                         &            \\
19.K2  & 73.5365975   & -3.0080161   &                         &            \\
19.K3  & 73.5368083   & -3.0077038   &                         &            \\
19.K4  & 73.5366495   & -3.0079314   &                         &            \\
\hline
20.1$\dagger$   & 73.5463230   & -3.0165205   & --             & New system \\
20.2$\dagger$   & 73.5456678   & -3.0159933   & --             &            \\
20.3$\dagger$   & 73.5450375   & -3.0154476   & --             &            \\
\enddata
\tablecomments{Column~(1): Lens system ID. Column~(2) \&~(3): R. A. and Dec. of lensed images. Column~(4): \texttt{EAZY}~\citep{2008ApJ...686.1503B} photometric redshift with 95\% confidence interval estimated using HST and JWST observations. Column~(5): Comments.}
\tablecomments{`$\star$': spectroscopic redshift taken from~\citet{2021MNRAS.508.1206J}.}
\tablecomments{`$\dagger$': lensed image candidate and not used in the lens reconstruction.}
\end{deluxetable}

\bibliographystyle{aasjournal}
\bibliography{reference}

\appendix
\section{Size estimates for knots in lensed arc at $z=6.70$ with model-1 and model-2.}

\begin{deluxetable}{ccccc}[!ht]
\label{tab:z6p70_app}
\tablecaption{Size measurements for knots at $\lowercase{z}=6.70$ with model-1 and model-2 discussed in \Sref{sec:swallow_6p7}.}
\tablewidth{\columnwidth}
\tablehead{
\colhead{ID}  & \colhead{$[\mu_t, \mu_r]$} & \colhead{$\theta_{\rm image}$} & \colhead{$\theta_{\rm source}$} & \colhead{$R_{\rm source}$} \\
              &                            & \colhead{(in mas)}       & \colhead{(in mas)}        & \colhead{(in pc)} }
\startdata
\multicolumn{5}{c}{model-1} \\
\hline
19.12  & [-25.3, 1.2]  & $<40.3$          & $<1.6$        & $<8.6$        \\
19.22  & [81.7, 1.2]  & $<30.6$          & $<0.4$        & $<2.0$         \\
19.32  & [742.8, 1.5] & $<32.7$          & $<0.1$        & $<0.2$         \\
19.K1  & [393.8, 1.3]  & $<30.2$          & $<0.1$        & $<0.4$         \\
19.K2  & [56.4, 1.4]   & $129.4 \pm 53.0$ & $2.3 \pm 0.9$ & $12.3 \pm 5.0$ \\
19.K3  & [150.3, 1.3]   & $<32.7$          & $<0.2$        & $<1.2$         \\
19.K4  & [60.7, 1.4]   & $<36.3$          & $<0.6$        & $<3.2$          \\
19.K5  & [54.8, 1.5]   & $78.9 \pm 24.4$  & $1.4 \pm 0.4$ & $7.7 \pm 2.4$  \\
\hline
\multicolumn{5}{c}{model-2} \\
\hline
19.12  & [-21.3, 1.2]  & $<40.3$          & $<1.9$        & $<10.1$        \\
19.22  & [184.5, 1.4]  & $<30.6$          & $<0.2$        & $<0.9$         \\
19.32  & [-293.8, 1.4] & $<32.7$          & $<0.1$        & $<0.6$         \\
19.K1  & [-517.5, 1.3]  & $<30.2$          & $<0.1$        & $<0.3$         \\
19.K2  & [96.4, 1.3]   & $129.4 \pm 53.0$ & $1.3 \pm 0.6$ & $7.2 \pm 3.0$ \\
19.K3  & [703.5, 1.3]   & $<32.7$          & $<0.1$        & $<0.2$         \\
19.K4  & [104.9, 1.3]   & $<36.3$          & $<0.3$        & $<1.9$          \\
19.K5  & [93.9, 1.4]   & $78.9 \pm 24.4$  & $0.8 \pm 0.3$ & $4.5 \pm 1.4$  \\
\enddata
\tablecomments{Column~(1): Knot system ID. Column~(2): Best-fit tangential and radial magnifications at the observed lensed images. Column~(3): $\theta_{\rm image}$ represents the measured/observed effective angular radius of the knot in milli-arcseconds. Column~(4): $\theta_{\rm source}$ represents the lensing corrected (i.e., observed size divided by tangential magnification) source plane effective angular radius of the knot in milli-arcseconds. Column~(5): $R_{\rm source}$ represents the lensing corrected source plane effective radius of the knot in parsecs. For a PSF-dominated source, we only provide upper bounds.}
\end{deluxetable}

\end{document}